# Hierarchical fringe tracker to co-phase and coherence very large optical interferometers

Romain G. Petrov[*,a], Abdelkarim Boskri[b,a], Yves Bresson[a], Karim Agabi[a], Jean-Pierre Folcher[a], Thami Elhalkouj[b], Stephane Lagarde[a], Zouhair Benkhaldoun [b].
[a]Université de la Côte d'Azur, Laboratoire Lagrange, UMR7293, OCA, 06304 Nice, France;
[b]Université Cadi-Ayyad, LPHEA, Marrakech, Maroc.

## ABSTRACT

The full scientific potential of the VLTI with its second generation instruments MATISSE and GRAVITY require fringe tracking up to magnitudes K>14 with the UTs and K>10 with the ATs. The GRAVITY fringe tracker (FT) will be limited to K~10.5 with UTs and K~7.5 with ATs, for fundamental conceptual reasons: the flux of each telescope is distributed among 3 cophasing pairs and then among 5 spectral channels for coherencing. To overcome this limit we propose a new FT concept, called Hierarchical Fringe Tracker (HFT) that cophase pairs of apertures with all the flux from two apertures and only one spectral channel. When the pair is cophased, most of the flux is transmitted as if it was produced by an unique single mode beam and then used to cophase pairs of pairs and then pairs of groups. At the deeper level, the flux is used in an optimized dispersed fringe device for coherencing. On the VLTI such a system allows a gain of about 3 magnitudes over the GRAVITY FT. On interferometers with more apertures such as CHARA (6 telescopes) or a future Planet Formation Imager (12 to 20 telescopes), the HFT would be even more decisive, as its performance does not decrease with the number of apertures. It would allow building a PFI reaching a coherent magnitude H~10 with 16 apertures with diameters smaller than 2 m. We present the HFT concept, the first steps of its feasibility demonstration from computer simulations and the optical design of a 4 telescopes HFT prototype.

## 1. INTRODUCTION

### 1.1. Cophasing and coherencing in optical interferometry

An astronomical optical interferometer combines the flux of several apertures to produce interference fringes. The contrast and relative position of these fringes yield the complex visibility of the source, that is the Fourier transform of its brightness distribution at the spatial frequencies $B/\lambda$, where B is the baseline (i.e. the distance between the apertures projected on the sky) and $\lambda$ the observation wavelength. The optical paths differences (OPDs) between the different apertures and the recombination focus must be reduced below a fraction of the coherence length $R\lambda$, where R is the spectral resolution of the focal instrument. On ground-based interferometers, the OPDs change very rapidly because of the atmosphere, and the fringes are affected by a fast jitter. A good sensitivity and accuracy of interferometric observations needs to stabilize the fringes by reducing the OPDs to a fraction of the observing wavelength to be able to integrate for a long time. This is called cophasing. When cophasing is impossible, but the OPDs can be kept smaller than the coherence length $R\lambda$, which is called coherencing, it is still possible to obtain information from a statistical analysis of many short exposures of the fringes. A device called a Fringe Tracker (FT) is used for cophasing and coherencing. The FT contains a Fringe Sensor (FS), a controller and actuators that compensate in real time the OPD variations. A high accuracy FT requires a FS with very short exposure times, typically a few ms in the near infrared. These short exposure times are the main reason for the sensitivity limit of FTs, and of optical interferometry in general.

### 1.2. Scientific specifications of a New Generation Fringe Tracker

On the VLTI we have demonstrated that the full potential of the second generation instruments GRAVITY[1] and MATISSE[2] can be achieved only if we can cophase at magnitudes K>14 with the UTs and K>10 with the ATs. With K>14 GRAVITY and MATISSE will be able to observe about 60 Quasars[3], and to constrain the geometry of their BLRs and inner dust torus. Combined with Reverberation Mapping (RM), this will validate and calibrate a method using RM alone or RM+spectro-astrometry to directly measure the masses and the distances of Quasars up to z>3. With K>9, the ATs would image a fair number of planet formation disks with a fair dynamical and temporal resolution. K>10 with the ATs would also very substantially improve the images of the brightest AGNs and their very complex dust environment. These magnitudes would also open many other programs, from binary asteroids to many topics in stellar physics. The GRAVITY FT that is currently in commissioning confirms that it will achieve about K=10.5 with UTs and K=7.5 with ATs. This is a major progress, but remains quite below the global VLTI needs. We will show below that the limitations of the GRAVITY FT are conceptual and that a new concept is necessary to achieve the desired sensitivity. Beyond the VLTI, a project like the Planet Formation Imager critically depends on the possibility to achieve a sensitivity of K>10 (in coherent flux) with modest 2 m class apertures.

## 1.3. Functions and accuracy of a fringe sensor

A fringe sensor has two fundamental functions. The main one is to measure the rapidly variable phase of the fringes (the *phase delay*) and stabilize it. If this is executed on nearly monochromatic fringes, there is a $2\pi$ ambiguity on the measurement and therefore a $\lambda$ ambiguity on the correction. Thus, phase delay measurement allows undetected fringe jumps. A fringe jump in the K band is a fraction of a fringe shift in the N band that will make coherent fringe integration impossible. Thus, the second function of a fringe tracker is to prevent fringe jumps or at the very least to detect them. This is performed by an analysis of the *group delay* of the whole fringe packet. In current systems, this is obtained by dispersing the light over $n_\lambda$ spectral channels. Then, the global $\lambda$ ambiguity is removed until the smallest common multiple of all channels wavelength. Coherencing is also a key feature for the robustness of the FT operation as it speeds up the acquisition and re-acquisition of fringes. There are several types of fringe sensors, with two main groups. In the all-in-one multi-axial devices, such as the instruments AMBER[4] and MATISSE[2], *all* apertures produce a common global interferogram where the different baselines are separated in data processing because they have different spatial frequencies. This system is mostly used when only a coherencing is desired. In pair-wise systems such as PIONIER[5] and GRAVITY[1], the differential piston (the average OPD between two apertures) is measured for *all* possible telescope pairs. This is the system most often used for cophasing. In both cases, the performance of the system decreases with the number of apertures. The two types of systems have similar performances, with a practical advantage for pair-wise cophasing systems, mainly because they allow shorter individual exposures. We will therefore use pair wise systems, and in particular GRAVITY, as reference values.

The two fundamental parameters of a combination of pair-wise fringe trackers are[6]:
- The number $n_{pair}$ of pair-wise FT fed by each aperture.
- The number $n_\lambda$ of spectral channels used for coherencing.

The accuracy $\sigma_{\varphi 1}$ of a phase delay measurement in a single spectral channel is given by[7] $\sigma_{\varphi 1}=1/(\sqrt{2}\,SNR_{C1})$ where $SNR_{C1}$ is the signal-to-noise ratio on the coherent flux in one spectral channel. If we consider only the source photon noise and the detector readout noise and neglect the thermal background in the near infrared to compute[8] the coherent flux SNR, we get:

$$\sigma_{\varphi 1} = \frac{1}{\sqrt{2}SNR_{C1}} \simeq \frac{\sqrt{2\frac{n_*}{n_\lambda n_{pair}}+n_{pix}\sigma_{RON}^2}}{\sqrt{2}V\frac{n_*}{n_\lambda n_{pair}}} \qquad (1)$$

If the phase accuracy per spectral channel is smaller than about 1 radian[7] ($\sigma_{\varphi 1}<1$), the measurement from all spectral channels can be combined to get the global phase delay accuracy $\sigma_\varphi$:

$$\sigma_\varphi = \frac{\sigma_{\varphi 1}}{\sqrt{n_\lambda}} \simeq \frac{\sqrt{2\frac{n_*}{n_\lambda n_{pair}}+n_{pix}\sigma_{RON}^2}}{\sqrt{2}V\frac{n_*}{n_\lambda n_{pair}}\sqrt{n_\lambda}} \qquad (2)$$

where
- $n_*$ is the number of coherent photons received from the source from each aperture.
- $n_{pix}$ is the number of pixels used for the measurement. Pairwise setups generally use the so-called "ABCD" approach that can be approximated by equation (2) with $n_{pix}=4$.
- $\sigma_{RON}$ is the standard deviation of the detector readout noise.
- $V$ is the instrument visibility (the source visibility affects the coherent flux $n_*$).

The necessity to have $\sigma_{\varphi 1}<1$ rad combined to the general specification on the piston accuracy $\sigma_p = \lambda\sigma_\varphi/2\pi < \lambda/p_{spec}$ yields the general condition for a cophaser:

$$\sigma_{\varphi 1} < min\left\{1, \frac{2\pi\sqrt{n_\lambda}}{p_{spec}}\right\} \quad (3)$$

Combining equations (1) and (3) and solving it in $n_*$ yields the minimum coherent flux:

$$n_* > n_{pair}.max\left\{n_\lambda\frac{1+\sqrt{1+2n_{pix}\sigma_{RON}^2 V^2}}{2V^2}, \frac{1+\sqrt{1+2n_\lambda n_{pix}\sigma_{RON}^2\left(\frac{2\pi}{p_{spec}}V\right)^2}}{2\left(\frac{2\pi}{p_{spec}}V\right)^2}\right\} \quad (4)$$

Except for high $p_{spec}$, the first term is larger and we can write[6]:

$$n_* > n_{pair}.n_\lambda \frac{1+\sqrt{1+2n_{pix}\sigma_{RON}^2 V^2}}{2V^2} = n_{pair}.n_\lambda.I_C \quad (4)$$

where $I_C$ is a characteristic of the cophaser that can be considered to be very similar for all systems using the same basic technology. It is therefore critical to reduce $n_{pair}$ and $n_\lambda$.

For example, the GRAVITY FT has $n_{pair} = 3$ and $n_\lambda = 5$. Here we discuss the possibility to build a system with $n_{pair} = 1$ and $n_\lambda = 1$, for any number of telescopes. If we succeed, our concept will have the highest possible sensitivity for a given technology level. For the VLTI with four apertures the potential gain in flux is 15, i.e. 3 magnitudes, if all other elements are the same, including the detector, the throughput, the exposure time and the control loop.

## 2. HIERACHICAL FRINGE TRACKING

### 2.1. The concept of hierarchical fringe tracking

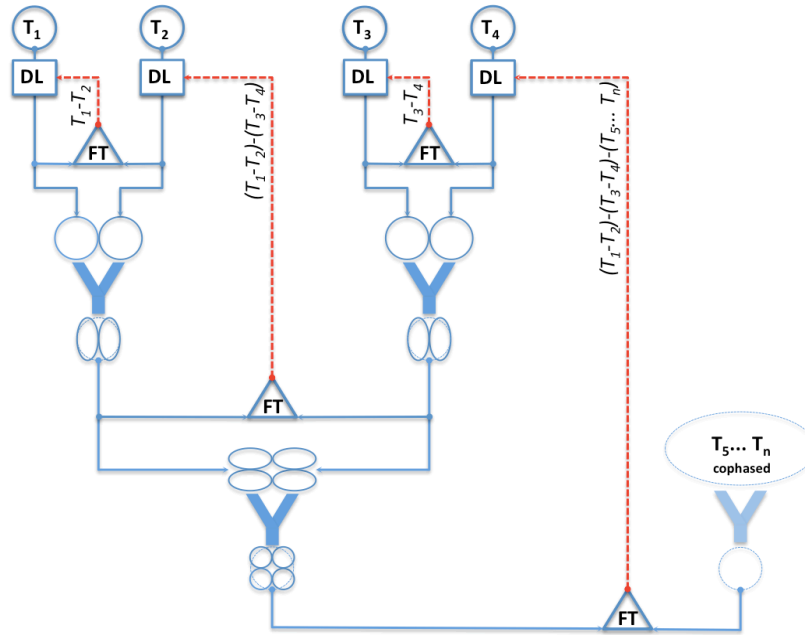

Figure 1: Concept of Hierarchical Fringe Tracking. We cophase pairs of telescopes, then pairs of pairs, then pairs of groups. Each fringe tracker FT sees two incoming beams, and leaves a fraction α>50% of the flux of each beam to cophase deeper level groups of telescopes. Each FT drives only one delay line (DL). There are solutions for any number of apertures but the best optimization can be achieved for $2^k$ telescopes. Then, all FTs receive $f_T 2^k/(2^k-1)$ photons where $f_T$ is the flux collected by one aperture.

The first idea of Hierarchical Fringe Tracking[10], illustrated in the figure 1, is to cophase the apertures by pairs, then by pairs of pairs, then by pairs of groups by a cascade of pair-wise FTs. Each FT drives one OPD actuator and it is easy to show that the necessary number of cophasers for $n_T$ telescopes is $n_T$-1. If at each level we transmit more than 50% of the flux to the deeper level, then the flux used by each fringe tracker will increase with the depth of the cophasing level. Such a system would be roughly equivalent to a $n_{pair} = 2$ system, whatever the number of apertures. There are other $n_{pair} = 2$ systems, such as a chain of FT where each aperture is cophased with only its two neighbors[9]. A possible advantage of the HFT is that the beams are at least partially transmitted even when a given FT is not working.

**2.2. Hierarchical Fringe Tracking with a 2 Beams Spatial Filter (TBSF)**

The concept of HFT could reach an ultimate level of performance if each pair-wise fringe tracker could:
- Transmit most (and in any case more than 50%) of the incoming flux to the deeper levels when the two input beams are cophased.
- Deflect some of the flux to measure the differential piston between the two beams, when they are not cophased.

Ideally, the output beam has all the characteristics of a single mode beam produced by a unique aperture. An easy way to understand such a system is to imagine that we transform the classical ABCD setup by merging two output channels in quadrature, to make them contain all phases included between -π/2 and π/2. We call this channel "C". When the input piston is 0, this combined channel C will contain 80% of the flux. "C" can be used to feed the next level of FT. The two remaining channels, "A" between -π and π/2 and "B" between π/2 and π, are in quadrature and can be used to measure the phase. For example (A-B)/(C-A-B) provides a phase estimator insensitive to changes in the flux ratio or in the source visibility.

Figure 2 illustrate this concept with integrated optics, by merging two of the outputs of a classical ABCD integrated optics beam combiner[11].

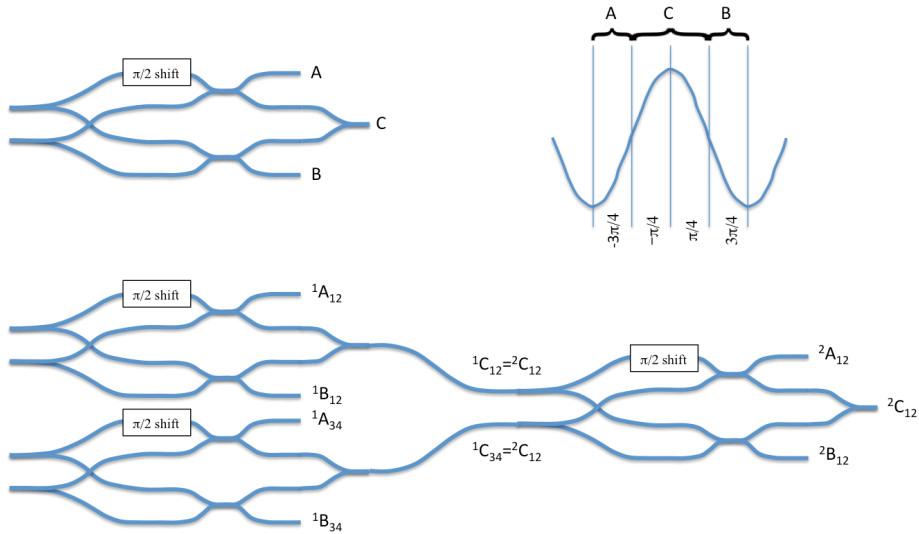

Figure 2: principle of an integrated optics TBSF. The upper left figure shows a classical ABCD integrated optics beam combiner where two outputs in quadrature are merged and used to feed the next level, as indicated by the upper right figure. The phase can be measured for example from (A-B)/(C-A-B). The combined output "C" can be used to feed another TBSF at the next cophasing level. This figure is intended to explain a concept.

We have not investigated the feasibility and the optimization of such an integrated optics device. We have started by simulations of a bulk optics achromatic TBSF, to better understand the different functions and their optimization.

## 3. A DISPERSED FRINGES BROADBAND ACHROMATIC 2 BEAMS SPATIAL FILTER

We are currently studying a possible implementation of a two beams spatial filter[10] (TBSF), with three additional requirements.
- The TBSF must be achromatic. We want the possibility to use the largest possible spectral band permitted by the source, the detector and the atmosphere. For example a FT system using all or most of the flux in the J, H

- and K band, where we have low noise detectors, good adaptive optics and still relatively slow turbulence, would be particularly interesting to support the observations of MATISSE in the L, M and N bands.
- We want to minimize the number of measures or pixels necessary for the piston measurement, as this sets the ultimate sensitivity limit of the 2T FT module.
- We would like to have an unambiguous piston measure in a range larger than the fringe. The ideal solution would be a measure in a range of about 10 μm, approaching the overall atmospheric piston excursion at relatively short time scales.
- A secondary objective is to use both polarizations of light without needing polarization correctors, which have been used with success at the VLTI with PIONIER but might be difficult to implement for very broad bands.

All these requirements suggested us to investigate a bulk optics very broadband solution based on dispersed fringes, which is described in figure 3. A pair on input pupils produces a set of dispersed Fizeau fringes in the focal plane of a lens. Here we have simulated very broadband fringes from 1.05 to 2.45 μm. As the piston changes, the fringes drift in all spectral channels at different speeds and the dispersed fringe figure "bents". In the focal plane, we put an intensity mask, built from a "photograph" of the dispersed fringes at piston=0. The bright fringes coincide with a transmission part and the dark fringes with a reflective part. The transmitted flux C goes through a sharp maximum when the piston is zero and displays strong minima around ±1μm, as illustrated by figure 4. The transmitted flux is anti-dispersed and collimated again. If the step of this so-called cophasing mask is equally divided between transmitting and reflecting parts, the peak flux of C at piston 0 will be of about 80%.

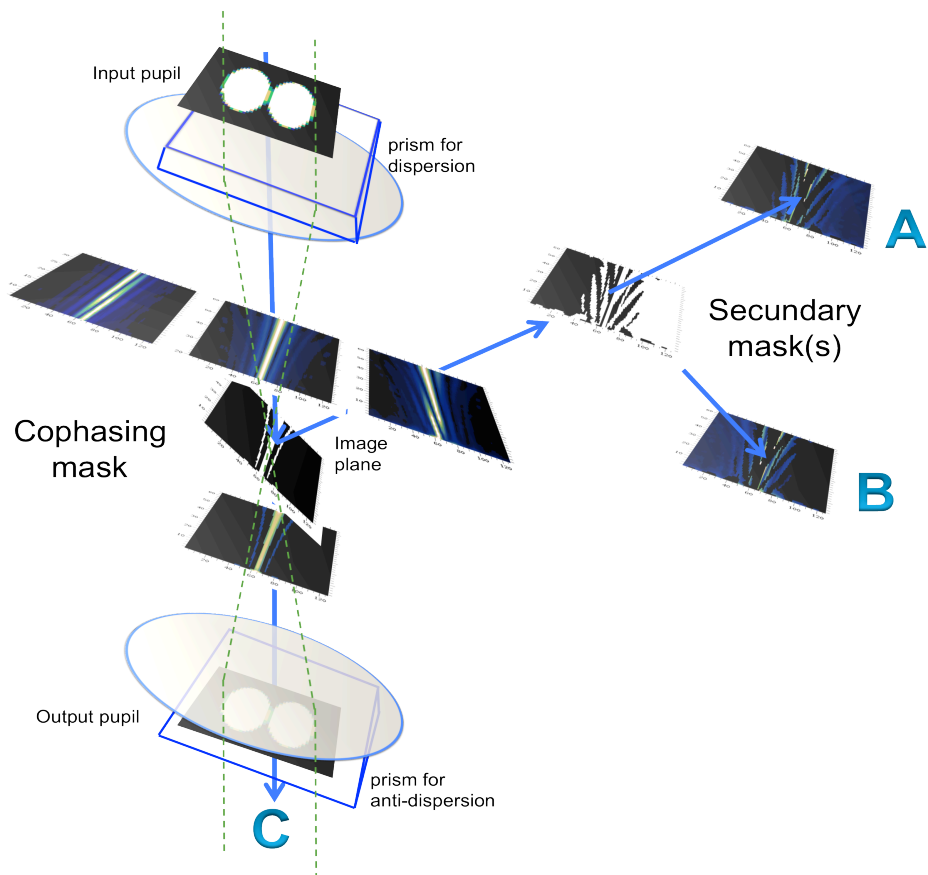

Figure 3: A broadband dispersed fringes achromatic 2 telescopes spatial filter and piston sensor.

When the input piston is non-zero, the bright fringes are reflected toward a second mask used to analyze the reflected pattern and estimate the piston. This second mask produces the fluxes "A" (transmitted) and "B" (reflected). Several masks are considered in order to maximize the "bijectivity zone", i.e. the range of pistons that are unambiguously obtained from the measures. Figure 4 displays the so-called "Marrakech solution", where the secondary mask is based on the fringes reflected by the first mask for a piston corresponding to the minimum of C. To maximize the dissymmetry between A and B, one half of the mask has been inverted, i.e. the reflecting parts became transmitting ones. This solution yields a quite large "bijectivity zone", of the order of 3 µm, but might not be optimized in terms of SNR for the small pistons. The definition of the optimum mask is a work in progress.

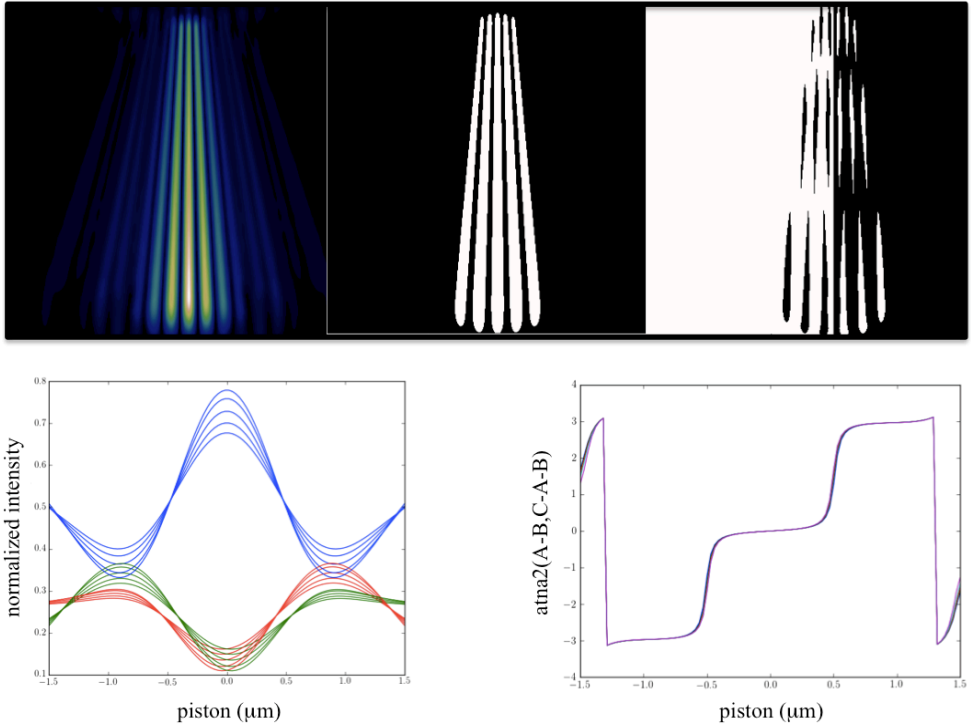

Figure 4: the "Marrakech solution" for the TBSF masks. Upper left: the dispersed fringes. Upper center: the primary mask transmitting "C". Upper right: the secondary mask transmitting "A" and reflecting "B". Lower left the signals A, B, and C as a function of the input piston, for flux ratios ranging from 3 to 1. Lower right atan2[(A-B)/(C-A-B)] as a function of the piston. This value is independent of the input flux ratio and source visibility (the curves for different flux ratios are superimposed). We have a fairly large zone where the piston can be recovered unambiguously, but the slope Estimator=f(piston) is small and the measure is too sensitive to noise.

To evaluate the quality of the spatial filtering we have simulated an input wavefront affected by standard Paranal perturbations and partially corrected to obtain a Strehl ratio of 0.5 on UTs. The results are displayed in figure 5 that shows that the output Strehl (and hence fringe contrast and piston noise) is quite insensitive to variations of the input wavefront and piston, with a maximum rms of 1.5% for a fairly large 2λ/D spatial filter.
Figure 6 shows the sensitivity of the piston variations transmitted by the TBSF to atmospheric noise. We see that the input piston is properly transmitted with a small multiplicative bias that could be calibrated but depends from the fringe contrast. We still have to check that this can be corrected using the contrast information given by the TBSF.

## 4. COHERENCING WITH THE HFT

The goal of a very broadband TBSF is to obtain an unambiguous piston measure in a large range with a single spectral channel. The optimization of this function is a work in progress but it seems already clear that the TBSF alone will not allow recovering pistons variations much larger than 2 µm. This is better than conventional narrow bandwidth ABCD systems, but not enough to avoid fringe losses in low flux regime. To complement the HFT, we can use the flux that is transmitted by the deepest cophasing level to feed an optimized x-λ group delay sensor. When a TBSF is far from cophased, the flux transmitted by each level is of the order of 50%, with a primary mask with equal transmitting and reflecting parts. This value can be increased, by making the transmitting steps larger, to the cost of a lower efficiency of the individual cophasers. With 4 telescopes, i.e. two levels, we have shown that the accuracy of a group delay sensor, optimized for coherencing, is comparable to the range of optimum TBSF cophasing for the limiting magnitude of HFT, if we assume a group delay cycle of 2 s, acceptable in terms of atmospheric piston variation (typically 1 μm variation per s with the ATs). With more apertures and hence levels, the matching between cophasing and last level coherencing might appear problematic. Then, we can extend the coherencing capacity of the TBSF by analyzing A and B in two spectral channels.

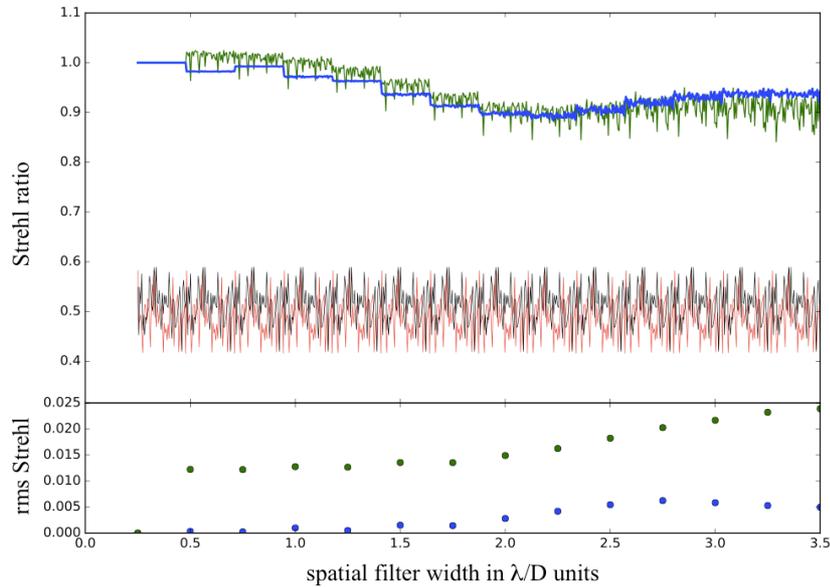

Figure 5. Spatial filtering quality of the achromatic TBSF. The black and red curves in the center show the instantaneous Strehl ratio on the two input pupils, with an average of 50%. The upper curves show the Strehl after the output pupils. The blue curve is for C maximum and the green one for C minimum. These plots are shown as a function of the spatial filter total size, in λ/D units. For each spatial filter size, we have displayed the Strehl ratios for 50 individual frames. The lower panel shows the rms of the output Strehl ratio variation. We see that output Strehl decreases down to 0.9 for a spatial filter of 2λ/D but is quite insensitive to variations of the input Strehl or piston, with a maximum rms of 1.5%

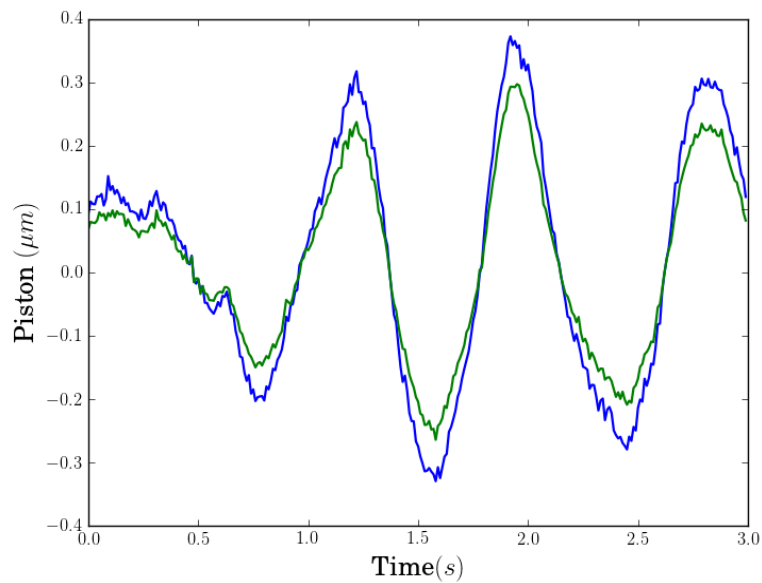

Figure 6: Sensitivity of the TBSF piston transmission to atmospheric noise. In blue the input piston variation affected by the atmospheric noise corresponding to a Strehl=0.5. In green the output piston variation. There is a small multiplicative bias that depends from the fringe contrast.

## 5. HFT PROTOTYPE AND TEST BENCH

Figure 7 shows the functions of a 4 apertures HFT. A source module contains OPD piezos actuator that can be used to introduce perturbation pistons with the statistical characteristics of atmospheric piston. A second set of piezos actuators will be used to correct the piston from the measures given by the HFT. The interferometric module contains the "C" beams and the main grids. Between the first and second level, the dispersion and grid directions are changed. Thus the first level grids are not seen in the second level images. The "C" beams are finally fed into a dispersed fringes group delay sensor. This device starts with a beam densifier designed to erase the image of the grids in the TBSF image planes: the pixel size of the group delay sensor is an integer multiple of the period of the grids. The photometric module produces the "A" and "B" beams and contains the secondary grids. This system is designed for each level to be "transparent" for the next one.

Figure 8 shows the principle of a compact optical setup to realize the TBSF displayed in figure 3. A unique collimator is used to produce all the images in the same plane containing the grids. To go from a grid to the next one, we use a 2f-2f mirror to reimage the "C" grid on the same focal plane, where we place also the "A-B" grid. In fact it is possible to use only one collimator for the 3 spatial filters as illustrated in the Zemax plot in figure 9.

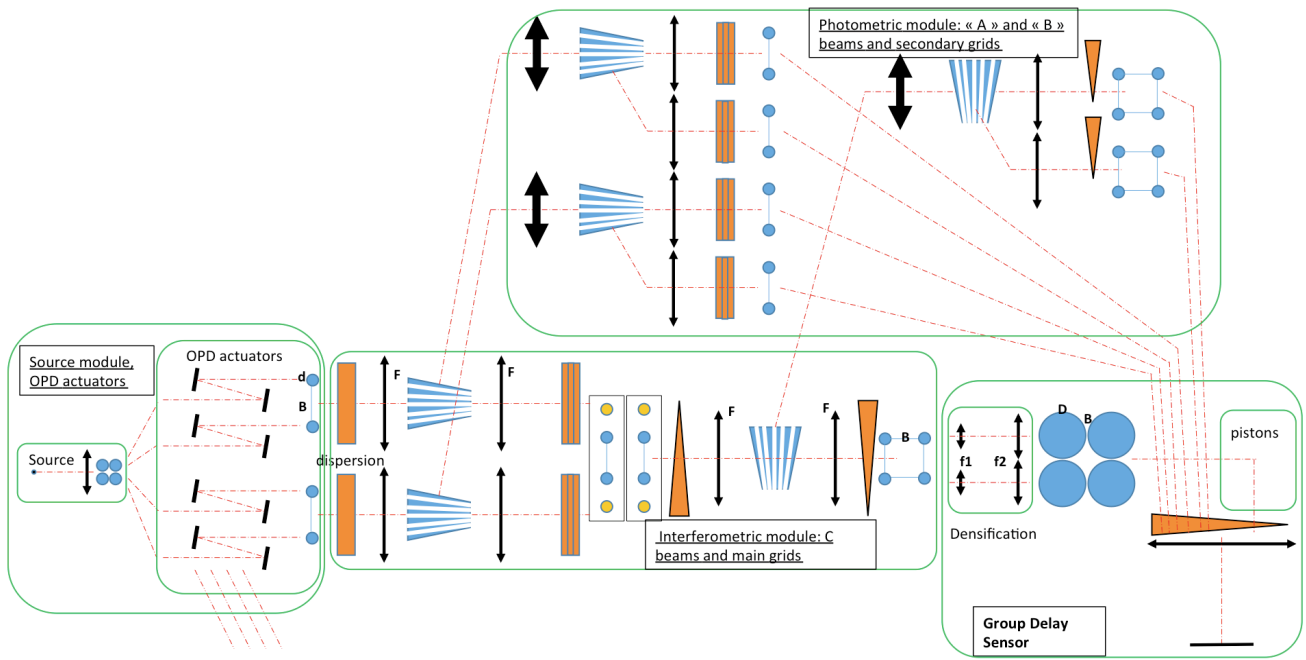

Figure 7: functional diagram of a 4 telescopes HFT test bench.

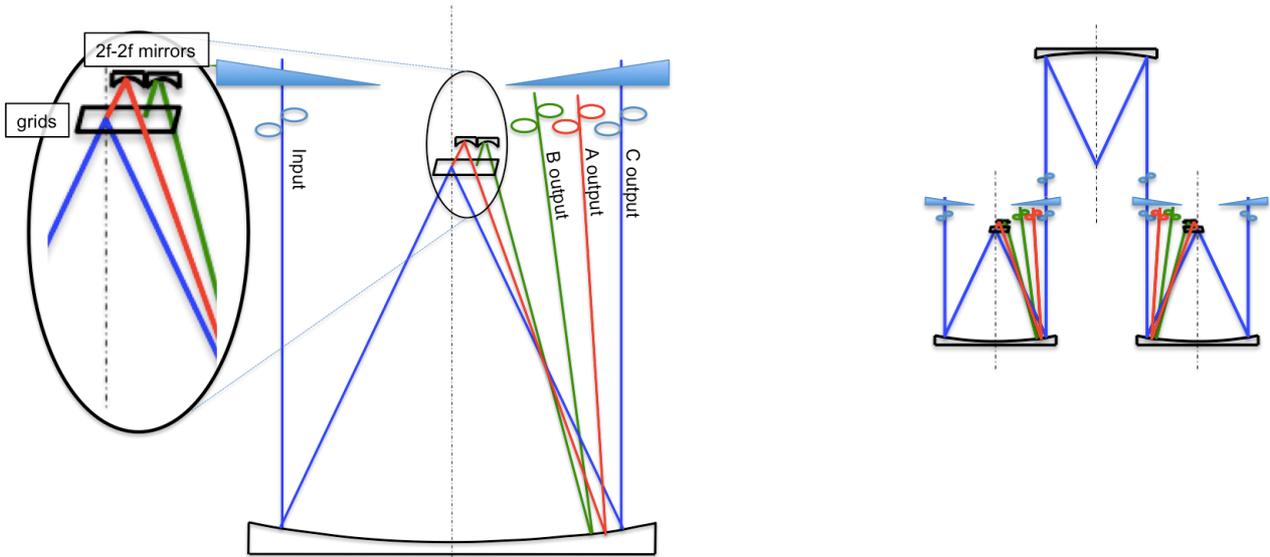

Figure 8: a compact implementation of a two beams spatial filter. The collimator produces the fringe image at his focus. In this setup "C" is reflected and the "C" beam is re-collimated by the collimator. The transmitted beam is re-imaged in the same plane by a 2f-2f mirror. Then, the second grid, which has been installed on the same support transmits "A" and reflects "B". A second 2f-2f mirror re-images "B" on the same focal plane, in order for the collimator to collimate the "B" beam as the "A" and "C" ones. In the upper right corner we have displayed a possible combination of TBSF to realize a 4T HFT. In fact the same collimator can be used for the 3 spatial filters, by sending the beams back toward the collimator.

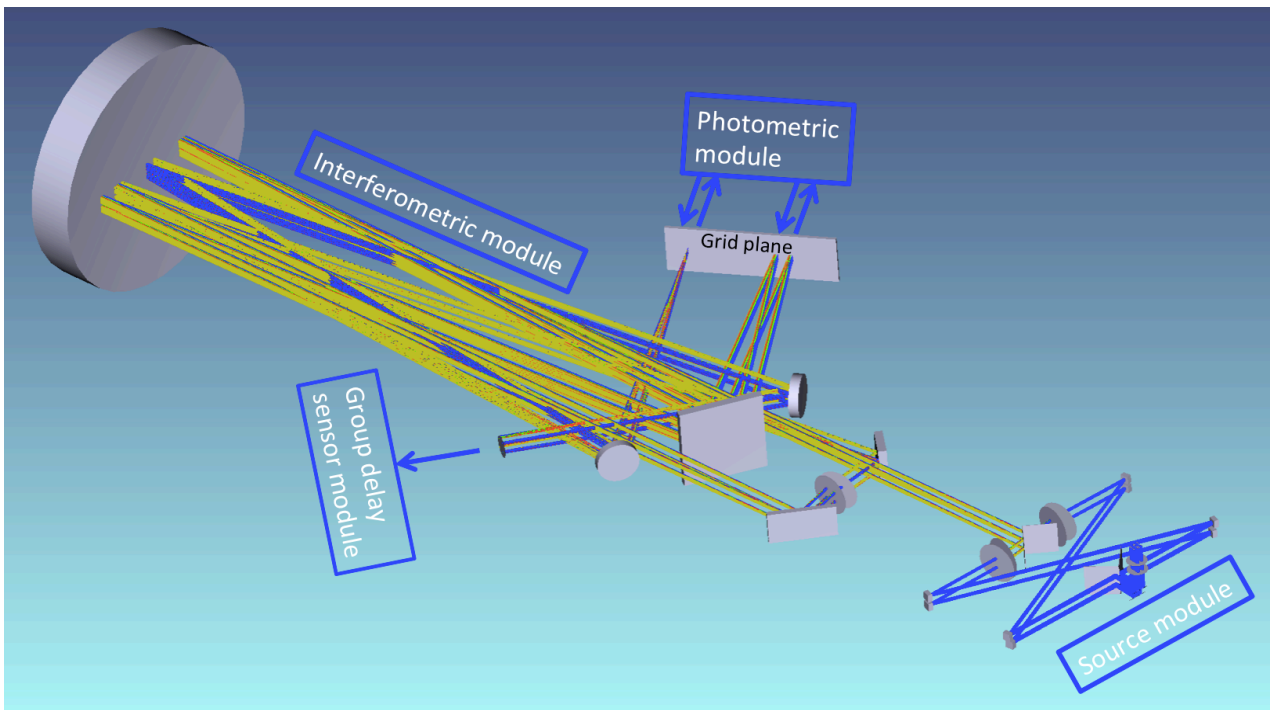

Figure 9: Zemax layout of the interferometric module of the HFT, with the source module. The input beams are parallel 2 by 2, but there is a small angle between the two pairs. Thus, they produce two different focal images in the grid plane, which contain the two grids of the first level TBSF. The re-collimated beams are sent back after a pair of mirrors making the two pairs of beams parallel. This produces the third focal image, common to all beams, and "C" grid in the same plane. The focal length of the collimator is 600 mm. The individual pupils have a diameter of 1.4 mm that yields a grid step of about 70 μm for an overall grid size of about 4 mm$^2$. The aperture ratios result in aberrations much smaller than the fringe and grid step.

## 6. CONCLUSION

The HFT is a system with $n_{pair}=1$ and $n_\lambda=1$ or $n_\lambda=2$. As such, it has the potential to improve the limiting sensitivity of the GRAVITY fringe Tracker by about 3 magnitudes. It would also allow building a 16 telescopes PFI with apertures substantially smaller than 2 m[6]. We are working on a full simulation of the HFT, including its control loop, to validate its performances and optimize its setup. This optimization includes the choice of the optimum grids, the optimum combination between cophasing by the two beams spatial filters and coherencing with the x-$\lambda$ group delay sensor. The prototype is intended to demonstrate the actual piston measures and to study the interactions between the different levels with real optics, even if the 4 telescopes HFT is designed for each level to be sensitive only to the global transmission of upper levels, without sensing their grids. If we can demonstrate that the HFT actually works, this device will quite certainly be the optimum FT for a given technological level and detector quality. Beyond the current bulk optics prototype with intensity masks, we will explore the possibility to realize an HFT with phase masks, which could make it easier for each level to be actually transparent for the next ones. We will also explore the possibility to design an HFT with integrated optics.